\documentclass[12pt, a4paper]{article}
\usepackage{amsfonts}
\usepackage{verbatim}
\usepackage{latexsym}
\usepackage{amsmath}

\begin{document}
\begin{flushright} arXiv: 0910.2880 [hep-th]\\ CAS-PHYS-BHU Preprint
\end{flushright} 

\vskip 1cm

\begin{center}
{\bf{\Large Rigid rotor as a toy model for Hodge theory}}

\vskip 2cm

{\bf Saurabh Gupta$^{(a)}$, R. P. Malik$^{(a, b)}$}\\
{\it $^{(a)}$Physics Department, Centre of Advanced Studies,}\\
{\it Banaras Hindu University, Varanasi - 221 005, (U.P.), India}\\

\vskip 0.1cm

{\bf and}\\

\vskip 0.1cm

{\it $^{(b)}$DST Centre for Interdisciplinary Mathematical Sciences,}\\
{\it Faculty of Science, Banaras Hindu University, Varanasi - 221 005, India}\\
{\small {\bf e-mails: guptasaurabh4u@gmail.com, malik@bhu.ac.in}}

\end{center}

\vskip 1.5cm

\noindent
{\bf Abstract:} We apply the superfield approach to the toy model of a rigid 
rotor and show the existence of the nilpotent and absolutely anticommuting
Becchi-Rouet-Stora-Tyutin (BRST) and anti-BRST symmetry transformations,
under which, the kinetic term and the action remain invariant. Furthermore, we also
derive the off-shell nilpotent and absolutely anticommuting (anti-) co-BRST symmetry
transformations, under which, the gauge-fixing term and the Lagrangian
remain invariant. The anticommutator of the above nilpotent symmetry transformations 
leads to the derivation of a bosonic symmetry transformation, under which, the
ghost terms and the action remain invariant. Together, the above transformations 
(and their corresponding generators) respect an algebra that turns out to be a 
physical realization of the algebra obeyed by the de Rham cohomological operators
of differential geometry. Thus, our present model is a toy model for the Hodge theory.\\

\vskip .1cm



\newpage

\noindent
{\bf \large 1. Introduction}\\
  
\noindent
The model of a rigid rotor, over the centuries, has played a pivotal role in providing
the key theoretical insights into the dynamics of the classical as well as quantum 
systems. In particular, in the realm of atomic, molecular and nuclear
physics, the contribution of the model of a rigid rotor has been enormous
(as far as deep understanding of many physical phenomena is concerned).
In our present investigation, we show that the above model 
presents a tractable toy model for the Hodge theory where the symmetry
transformations (and their corresponding generators) provide a physical
realization of the de Rham cohomological operators of differential geometry.

In our present endeavor, we begin with an {\it appropriate} first-order 
Lagrangian (FOL) for a particle of mass $m = 1$ that is constrained to move on a 
circle of radius $a$ (i.e. the model of a rigid rotor). This Lagrangian
is given below\footnote{We slightly differ from the FOL of [1] for the sake of brevity and
algebraic convenience.} (see, e.g. [1] for a detailed discussion on its appropriateness)
\begin{eqnarray}
L_f = p_r \; \dot r + p_\theta \; \dot \theta - \frac{p_\theta^2}{2 r ^2} 
+ \lambda \; (r - a),
\end{eqnarray}
where $ r $ and $ \theta $ are the polar coordinates, $\dot r = (dr/dt), 
\dot \theta = (d \theta /dt)$ are the generalized velocities, $p_r$ and 
$p_\theta$ are the corresponding canonical momenta and $\lambda$ is the Lagrange
multiplier. All these variables are function of the evolution parameter $ t $.
It can be seen that $\Pi_\lambda \approx  0$ and $(r - a) \approx 0$
are the two first-class constraints of the theory [1] where $\Pi_\lambda$ is the 
momentum corresponding to the Lagrange multiplier variable $\lambda$. The 
existence of the first-class constraint, as is well-known [2,3], is the signature of a 
gauge theory. In fact, the following infinitesimal local gauge 
transformations ($\delta_g$) [1]
\begin{eqnarray}
\delta_g p_r = f(t), \qquad \delta_g \lambda = \dot f(t), 
\qquad \delta_g r = \delta_g \theta = \delta_g p_\theta = 0,
\end{eqnarray}
are generated by the above first-class constraints, under which, we can
verify that the FOL transforms (with the infinitesimal gauge parameter $f(t)$) as
\begin{eqnarray} 
\delta_g L_f \; = \;  \frac{d}{dt} \; \Bigl (f(t) \; \bigl [r(t) - a \bigr ] \Bigr ). 
\end{eqnarray}
Thus, the action $ S = \int (dt\; L_f)$, corresponding to above Lagrangian, remains 
invariant under the infinitesimal local gauge symmetry transformations (2). The
conserved charge, that emerges from the above symmetry transformations
(due to the Noether's theorem), is nothing but the constraint $(r - a)$.

One of the most intuitive approaches to quantize a gauge theory is the 
Becchi-Rouet-Stora-Tyutin  (BRST) formalism where the gauge 
parameter is replaced by (anti-) ghost fields. We take up the 
Lagrangian (1) and demonstrate, in our present paper, that its (anti-) 
BRST invariant version (cf. (4) below) is endowed with 
a set of six continuous symmetry transformations which act 
infinitesimally on the variables of the theory.    
In fact, we shall establish that the algebra of the continuous symmetry 
operators is exactly same as the algebra obeyed by the 
cohomological operators of differential geometry. 
We also demonstrate that the latter algebra is 
respected by the conserved charges that generate the above continuous 
symmetry transformations. Thus, we prove that the rigid rotor
is a toy model for the Hodge theory where all the cohomological operators 
of differential geometry are identified with the continuous symmetries 
and their corresponding generators.

A couple of mathematical properties, that are associated with 
the key concepts of BRST formalism,
are the nilpotency and the absolute anticommutativity of the (anti-) BRST symmetry
transformations (and their corresponding generators). The geometrical origin and
interpretation of the above properties are provided by the superfield 
formalism [4,5] where the horizontality condition (HC) plays a decisive role.
The latter condition (i.e. (HC)) physically implies that the curvature tensor 
of a given gauge theory is unaffected by the presence of the Grassmannian 
variables invoked in the superfield description of the BRST formalism [4,5]. 
The (anti-) BRST transformations, for the gauge field and (anti-)
ghost fields, are determined by utilizing the HC within the framework
of superfield formalism. The components of the ``gauge" potential of the rigid rotor 
possess some unusual properties. For instance, the components of this 
potential transform in a completely different manner (cf. (2),(5),(6)). To obtain
such type of unusual transformations, within the framework of the 
superfield formulation, is a challenging problem for us. We have accomplished 
this goal of obtaining the  (anti-) BRST transformations for the ``gauge" components and 
(anti-) ghost variables of the rigid rotor by making some specific choices
in the application of the HC.

The main motivations for carrying out our present investigations are as
follows. First, we have already proposed, in our earlier works [6-10], the 
{\it field theoretic} models for the Hodge theory in the case of 2D (non-) 
Abelian 1-form and 4D Abelian 2-form gauge theories. Thus, it is an interesting
task for us to propose a {\it simple} toy model for the Hodge theory
where there are almost no mathematical complications and the continuous 
symmetry transformations of the theory are found to be completely transparent. 
Second, for the present toy model, the components of the ``gauge'' potential
transform in a completely different manner (see, e.g. [1]). It is,
therefore, an interesting endeavor to exploit the superfield formalism to obtain
such kind of symmetries within its geometrical framework. Finally, a new
model for the Hodge theory is always an exciting development
because its proof requires a variety of continuous symmetry
transformations. The ensuing operator algebra of the above
continuous symmetry transformations turns out to be reminiscent of the
algebra obeyed by the de Rham cohomological operators of differential
geometry. In other words, we establish a {\it perfect} analogy between
some aspects of mathematics of the differential geometry and the continuous symmetries of the
Lagrangian of our present toy model of a rigid rotor.

The contents of our present paper are organized as follows. In section 2, 
we discuss the off-shell nilpotent and absolutely anticommuting (anti-) BRST 
symmetry transformations within the frameworks of 
Lagrangian formalism and superfield approach. This is followed,
in section 3, by our discussion about the 
nilpotent and absolutely anticommuting dual(co)-BRST and anti-co-BRST 
symmetry transformations in the realm of Lagrangian formulation. 
Our subsequent section 4 deals with the derivation of a bosonic
symmetry transformation. In section 5, we discuss the ghost and discrete symmetry
transformations in the theory.
We demonstrate, in section 6, the similarity between
the algebra of the de Rham cohomological operators and continuous symmetry 
transformations (and corresponding generators). Finally, we make some concluding
remarks and point out a few future directions in section 7.

In our Appendix, we capture some of the key mathematical properties of the (anti-) BRST
charges in the superfield formalism. \\

\noindent
{\bf \large 2. Nilpotent and absolutely anticommuting (anti-)BRST

symmetries: two approaches}\\

\noindent
In this section, we discuss about the off-shell nilpotent (anti-) BRST
symmetries in the Lagrangian as well as the superfield formulation. First,
in subsection 2.1, we dwell on the completely different type of transformations
associated with the components of the ``gauge'' potential. Later on, in subsection 2.2, we
capture these unusual bit of transformations within the framework of 
the geometrical superfield formalism [4,5] with judicious choices.\\

\noindent
{\bf \large 2.1 (Anti-) BRST symmetries: Lagrangian description}\\

\noindent
We begin with the following (anti-) BRST invariant first-order appropriate 
Lagrangian for the rigid rotor (see, e.g. [1] for details)
\begin{eqnarray}
L_b = p_r \;\dot r + p_\theta \;\dot \theta - \frac{p^2 _\theta}{2 r^2} + \lambda \; (r - a)
+ b \; (\dot \lambda - p_r) + \frac {b^2} {2} - i \; \dot {\bar C}\; \dot C - i \; \bar C\; C,
\end{eqnarray}
which is the generalization of the first-order gauge-invariant Lagrangian (1). 
In the above, $\lambda$ and $p_r$ are the analogues
of the components of the ``gauge" potential and $b$ is the Nakanishi-Lautrup type 
auxiliary field that is invoked for the linearization of the gauge-fixing
term $ [- \frac{1}{2} (\dot \lambda - p_r)^2]$. The fermionic (anti-) ghost
fields $(\bar C) C$ are required for the validity of unitarity.

The above FOL respects the following infinitesimal and continuous (anti-) BRST symmetry
transformations $ s_{(a)b}$ (see, e.g. [1])
\begin{eqnarray}
& s_{ab}\; \lambda = \dot{\bar C}, \quad s_{ab} \;\bar C = 0, \quad s_{ab} \;p_r = \bar C,
\quad s_{ab} \;C = - i b, & \nonumber\\
& s_{ab}\; b = 0, \quad s_{ab} \;r = 0, \quad s_{ab} \;\theta = 0, \quad s_{ab} \;p_\theta = 0,& 
\end{eqnarray}
\begin{eqnarray}
& s_{b} \;\lambda = \dot{C}, \quad s_{b} \;C = 0, \quad s_{b} \;p_r = C,
\quad s_{b} \;\bar C = + i b, & \nonumber\\
& s_{b} \;b = 0, \quad s_{b} \;r = 0, \quad s_b \;\theta = 0, \quad s_{b}\; p_\theta = 0.& 
\end{eqnarray}
It can be seen that the following statements are true, namely;

(i) operator equations $s_b^2 = 0, s_{ab}^2 = 0, s_b s_{ab} + s_{ab} s_b = 0 $
are always satisfied,

(ii) out of the two momenta $p_r$ and $p_\theta$, only one of them transforms
(i.e. $s_{(a)b} p_r \not = 0, s_{(a)b} p_\theta = 0$) under the (anti-) BRST
symmetry transformations,

(iii) the coordinates and their velocities do not transform (i.e. $s_{(a)b} r 
= s_{(a)b} \theta = 0, \; s_{(a)b} \dot r =  s_{(a)b} \dot \theta = 0$) under
the (anti-) BRST transformations. As a result, the kinetic term of the theory
remains invariant, and

(iv) the transformation property of $\lambda$ and $p_r$ are quite different.\\
\noindent
The above observations imply that the gauge and (anti-) BRST invariant quantities 
$ r, \theta, p_\theta,$ etc., are ``physical" quantities (in some sense). These 
comments would play very important roles in our superfield approach to BRST formalism
(where we shall derive all the nilpotent and anticommuting
(anti-) BRST transformations for the all the variables
of the theory).

The conserved and nilpotent (anti-) BRST charges $Q_{(a)b}$,
that are the generators of the above (anti-) BRST symmetry transformations, are 
\begin{eqnarray}
Q_b = b \; \dot C - \dot b \; C, \qquad Q_{ab} =   b\;  \dot{\bar C} 
- \dot b \; \bar C.  
\end{eqnarray}
Using the following equations of motion 
\begin{eqnarray}
& \dot r = b, \qquad \dot b = r-a, \qquad {\ddot C} - C =0, \qquad \ddot {\bar C} 
- \bar C = 0, & \nonumber\\
& \dot p_\theta = 0, \qquad \dot p_r = \lambda + r \; \dot \theta^2,
 \qquad p_\theta = r^2 \; \dot \theta,& 
\end{eqnarray} 
it is straightforward to check that $\dot Q_b = 0, \; \dot Q_{ab} = 0$.
Furthermore, using the definition of the following canonical momenta 
\begin{eqnarray}
\Pi_\lambda = b, \qquad \Pi_C = + i \; \dot{\bar C}, \qquad \Pi_{\bar C} = - i \; \dot C,
\qquad \Pi_r =\; p_r, \qquad \Pi_{\theta} = \;p_\theta, 
\end{eqnarray}
and the  corresponding 
canonical brackets (cf. (44) below), it can be checked that the conserved charges obey
$Q_b^2 = 0, \; Q_{ab}^2 = 0, \; Q_b Q_{ab} + Q_{ab} Q_b = 0 $.

Under the (anti-) BRST symmetry transformations, the kinetic term 
$ p_\theta \dot\theta - p_\theta^2 / 2 r^2  \equiv r^2 \dot \theta^2/ 2$ 
remains invariant because $s_{(a)b} \; r = 0, \; s_{(a)b} \; \theta = 0, 
\; s_{(a)b} \; p_\theta = 0$. This kinetic term is equivalent to ($v^2/2$) 
where $ v = r \dot \theta$ is the linear velocity. In the language of 
differential geometry, the kinetic term owes its origin to the exterior 
derivative $ d = dt \; \partial_t$ (with $d^2 = 0$) 
because $ d x = dt \;(\dot x) \equiv dt\; (v_x), \; d y = dt\; (\dot y) 
\equiv dt \;(v_y),\;  v^2 = v_x^2 + v_y^2$. Thus, the nilpotent (anti-) BRST symmetry
transformations can be identified with the exterior derivative. The absolute 
anticommutativity of the (anti-) BRST transformations, however, imply that only
one of them can be identified with the exterior derivative $ d = dt \; \partial_t$
(with $d^2 = 0$) of the differential geometry.

We close this subsection with the remark that the physicality
condition $Q_{(a)b} |phys> = 0$ leads to the requirement that $ b |phys> = 0$
and $\dot b |phys> = 0$. These conditions, due to equations (8) and (9),
imply that the operator form of the first-class constraints $\Pi_\lambda \approx 0$
and   $(r - a) \approx 0$ annihilate the physical states of the theory. Thus,
the physicality criteria $Q_{(a)b} |phys> = 0$
is consistent with the Dirac's method of quantization
(see, e.g. [3] for details).   \\

\noindent
{\bf \large 2.2 (Anti-) BRST symmetries: superfield formalism}\\

\noindent
It is evident from our earlier discussions (cf. Introduction) that, out of two
polar coordinates, only one is independent coordinate parameter because of the
constraint condition $ r - a \approx 0$. Thus, effectively, the dynamics of the 
rotor will be described in terms of the space-time coordinates\footnote 
{Only in the framework of the superfield formulation [4,5], we shall take these
coordinates as independent variables. Ultimately, however, we shall take the limit
$r \to 0$ so that all the variables of the theory become function of the
evolution parameter $t$ only.}
$(r, t)$. As a consequence, we can define an exterior derivative (see, Sec. 7 below)
\begin{eqnarray}
d = dt\; \partial_t + dr \;\partial_r, \quad dr \wedge dt = - dt \wedge dr, \quad
dt \wedge dt = dr \wedge dr = 0.
\end{eqnarray}
Further, it is clear from the first-class constraints $\Pi_{\lambda} \approx 0$
and $(r - a) \approx 0 $ that the rigid rotor is a model of a gauge theory with 
gauge potentials represented by variables $\lambda$ and $p_r$. We can define a 
1-from connection as 
\begin{eqnarray}
A^{(1)} \; = \; dt \; \lambda (r, t ) + dr \; B (r, t),
\end{eqnarray}
where the gauge potential component $B(r, t)$ would be connected (as we shall 
see later) with $p_r$ in an explicit fashion. The curvature 2-from 
\begin{eqnarray}
d A^{(1)} \; = \;  (dt \wedge dr) \; [\partial_t B(r, t) - \partial_r \lambda(r, t)],
\end{eqnarray}
would remain invariant under the gauge (or (anti-) BRST) transformation.

In the superfield approach to BRST formalism (see, e.g. [4,5]), the exterior derivative and 1-form
connection are generalized to a supermanifold parametrized by $r, t, \eta$ 
and $\bar \eta$ where $\eta$ and $\bar\eta$ are Grassmannian variables (i.e.
$\eta^2 = \bar\eta^2 = 0, \eta \bar\eta + \bar\eta \eta = 0 $). Thus, 
on the above (2, 2)-dimensional supermanifold, we have the following expressions
for the generalizations, namely;
\begin{eqnarray}
d \rightarrow \tilde d &=& dt \; \partial_t + dr \; \partial_r + d \eta \; \partial_{\eta} 
+ d \bar\eta \; \partial_{\bar\eta}, \qquad \tilde d^2 = 0, \nonumber\\
A^{(1)} \rightarrow \tilde A^{(1)} &=& dt \; \tilde {\lambda} (r, t, \eta, \bar \eta) 
+ dr \; \tilde B (r, t, \eta, \bar \eta) \nonumber\\
&+& d \eta \; \bar F (r, t, \eta, \bar \eta) 
+ d \bar \eta \; F (r, t, \eta, \bar \eta),
\end{eqnarray}
where $\partial_{\eta} = (\partial/ \partial\eta),  \partial_{\bar \eta}
= (\partial / \partial\bar\eta)$ are the Grassmannian derivatives and 
the component superfields 
can be expanded along the Grassmannian directions as:
\begin{eqnarray} 
&& \tilde \lambda (r, t, \eta, \bar \eta) = \lambda (r, t) + \eta \; \bar f_1 (r, t)
+ \bar \eta \; f_1 (r, t) + i \; \eta \; \bar\eta \; B_1 (r, t), \nonumber\\
&& \tilde B (r, t, \eta, \bar \eta) = B (r, t) + \eta \; \bar f_2 (r, t)
+ \bar \eta \; f_2 (r, t) + i \; \eta \; \bar\eta \; B_2 (r, t), \nonumber\\
&& F (r, t, \eta, \bar \eta) = C(r, t) + i\; \eta \; \bar b_1 (r, t)
+ i\; \bar \eta \; b_1 (r, t) + i \; \eta \; \bar\eta \; S (r, t), \nonumber\\
&& \bar F (r, t, \eta, \bar \eta) = \bar C (r, t) + i \;\eta \; \bar b_2 (r, t)
+ i\; \bar \eta \; b_2 (r, t) + i \; \eta \; \bar\eta \; \bar S (r, t),  
\end{eqnarray}
where $(\bar C) C$ are the (anti-) ghost fields, $\bar f_1, f_1, \bar f_2, f_2, S, 
\bar S$ are the fermionic secondary fields and $ B_1, B_2, b_1, \bar b_1, b_2,
\bar b_2$ are the bosonic secondary fields that would be determined in terms of the
basic fields by exploiting the so-called horizontality condition of the geometrical
superfield formalism.

The celebrated horizontality condition requires that the gauge invariant curvature
2-form should remain independent of the Grassmannian variables $\eta$ and $\bar
\eta$. This can be expressed, in the mathematcal form, as 
\begin{eqnarray} 
\tilde d \tilde A^{(1)} = d A^{(1)}.                              
\end{eqnarray}
The l.h.s. of the above condition can be explicitly expressed as
\begin{eqnarray}
\tilde d \tilde A^{(1)} &=& (dt \wedge dr) [\partial_t \tilde B 
- \partial_r \tilde \lambda] + (dt \wedge d \eta) [\partial_t \bar F 
- \partial_{\eta} \tilde \lambda] + (dt \wedge d \bar \eta) 
[\partial_t F - \partial_{\bar \eta} \tilde \lambda] \nonumber\\
&+& (dr \wedge d \eta) [\partial_r \bar F - \partial_{\eta} \tilde B] 
+( dr \wedge d \bar \eta ) [\partial_r F - \partial_{\bar\eta} \tilde B]
- (d \eta \wedge d \eta ) (\partial_{\eta} \bar F) \nonumber\\
&-& (d \bar \eta \wedge d \bar \eta ) (\partial_{\bar \eta} F)
- (d \eta \wedge d \bar \eta ) [\partial_{\eta} F 
+ \partial_{\bar\eta} \bar F].
\end{eqnarray}
Ultimately, the horizontality condition (i.e. $\tilde d \tilde A^{(1)} 
= d A^{(1)}$) yields the following relationships between the basic and 
secondary fields:
\begin{eqnarray}
&& b_1 = 0, \qquad  \quad \bar b_2 = 0, \qquad \quad S = 0, \qquad \quad \bar S =0, 
 \nonumber\\
&& \bar f_1 = \dot{\bar C}, \qquad f_1 = \dot C, \qquad B_1 = \dot b_2 = - \dot {\bar b}_1, 
\qquad \bar f_2 = \partial_r \bar C, \nonumber\\ 
&& f_2 = \partial_r C, \qquad B_2 =\partial_r b_2 \equiv - \partial_r \bar b_1,
\qquad \bar b_1 + b_2 = 0.
\end{eqnarray}
The choice $b_2 = b =-\bar b_1$, yields the following explicit expressions
\begin{eqnarray}
&& \tilde \lambda ^{(R)}(r, t, \eta, \bar \eta) = \lambda (r, t) + \eta \; \dot {\bar C} (r, t)
+ \bar \eta \; \dot C (r, t)+ i \;  \eta \; \bar\eta \; \dot b(r, t), \nonumber\\ 
&& \tilde B ^{(R)} (r, t, \eta, \bar \eta) = B (r, t) + \eta \;  \partial_r \bar C (r, t)
+ \bar \eta \; \partial_r C (r, t) + i \; \eta \; \bar\eta 
\; \partial_r b (r, t), \nonumber\\
&& F ^{(R)} (r, t, \eta, \bar \eta) = C(r, t) - i \; \eta \; b (r, t), \nonumber\\
&& \bar F ^{(R)}  (r, t, \eta, \bar \eta) = \bar C (r, t) + i \; \bar \eta \; b (r, t),  
\end{eqnarray}
where the superscript (R) denotes the reduced form of the expansions in (14).
The stage is now set to make a judicious choice of $\tilde B^{(R)} (r, t, \eta, \bar\eta)$
and $ B(r, t)$ in terms of the gauge components $p_r(r, t)$. The following choices
\begin{eqnarray}
\tilde B^{(R)} (r, t, \eta, \bar\eta) = \partial_r \tilde p_r^{(R)} (r, t, \eta, \bar\eta), 
\qquad B (r, t) = \partial_r p_r(r, t),
\end{eqnarray}
lead to the following expansion for the super dynamical field: 
\begin{eqnarray}
\tilde p_r^{(R)} (r, t, \eta, \bar\eta) = p_r (r, t) + \eta \; \bar C(r, t)
+ \bar \eta \;  C(r, t) + i \; \eta \; \bar\eta \; b (r, t).
\end{eqnarray}
It is clear from the beginning that all the variables of our present toy
model are function of the evolution parameter $ t $ only. Thus, at this
juncture, we take the limit $ r \to 0 $, so that we obtain the following
correct physical expansion (corresponding to (18)) for the super dynamical
variables,  namely;
\begin{eqnarray}
&& \tilde \lambda ^{(h)} (t, \eta, \bar \eta) = \lambda (t) + \eta \; \dot {\bar C} (t)
+ \bar \eta \; \dot C (t)+ i \;  \eta \; \bar\eta \; \dot b(t), \nonumber\\ 
&& \tilde p_r ^{(h)} (t, \eta, \bar \eta) = p_r (t) + \eta \;  \bar C (t)
+ \bar \eta \;  C (t) + i \; \eta \; \bar\eta \;  b (t), \nonumber\\
&& F ^{(h)} (t, \eta, \bar \eta) = C(t) - i \; \eta \; b (t), \nonumber\\
&& \bar F ^{(h)} (t, \eta, \bar \eta) = \bar C (t) + i \; \bar \eta \; b (t),  
\end{eqnarray}
where the superscript (h) denotes the superfield expansions after the application
of the HC. In terms of the (anti-) BRST symmetry transformations (cf. (5), (6)), we have 
the following uniform expansions  
\begin{eqnarray} 
&& \tilde \lambda ^{(h)} (t, \eta, \bar \eta ) = \lambda  + \eta \; (s_{ab} \; \lambda ) 
+ \bar \eta \; (s_b \; \lambda ) +  \eta \; \bar\eta \; (s_b \; s_{ab} \; \lambda ), \nonumber\\
&& \tilde p_r ^{(h)} (t, \eta, \bar \eta) = p_r + \eta \; (s_{ab} \; p_r)
+ \bar \eta \; (s_b \; p_r) +  \eta \; \bar\eta \; (s_b \; s_{ab} \; p_r), \nonumber\\
&& F ^{(h)} (t, \eta, \bar \eta) = C + \eta \; (s_{ab} \; C) 
+ \bar \eta \; (s_b \; C) +   \eta \; \bar\eta \; (s_b \; s_{ab} \; C), \nonumber\\
&& \bar F ^{(h)} (t, \eta, \bar \eta) = \bar C + \eta \; (s_{ab} \; \bar C)
+ \bar \eta \; (s_b \; \bar C)+  \eta \; \bar\eta \; (s_b \; s_{ab} \; \bar C),  
\end{eqnarray}  
where we have taken into account the nilpotent symmetry transformations 
$s_b C =0, \; s_{ab} \bar C = 0, \; 
s_b s_{ab} C = s_b (- i b) = 0, \; s_b s_{ab} \bar C= 0$.

This establishes the non-trivial (anti-) BRST symmetry transformations of the 
dynamical variables of the theory. The trivial transformations $s_{(a)b} r = 0, \;
s_{(a)b} \theta = 0, \; s_{(a)b} p_{\theta} = 0$ can also be derived by exploiting the 
augmented superfield formulation [11-14] in which, in addition to the HC, 
the gauge (BRST) invariant quantities, too, are required to be independent of the Grassmannian   
variables. In other words, we demand the following\footnote {All the super dynamical 
variables, unlike in the context of the HC, would now be taken to be 
function of the (1, 2)-dimensional superspace variables $ t, \eta, \bar \eta$ only.} 
\begin{eqnarray}
\tilde r (t, \eta, \bar\eta) = r (t), \qquad
 {\tilde \theta} (t, \eta, \bar\eta) =  \theta (t), \qquad
\tilde p_\theta (t, \eta, \bar\eta) = p_\theta (t),\label{sup2}
\end{eqnarray}
where we have exploited the fact that the gauge invariant (physical) quantities are unaffected by
the presence of the Grassmannian variables within the framework of the augmented superfield formalism [11-14].
The following explicit expansions along the Grassmannian directions
\begin{eqnarray} 
&& \tilde r ( t, \eta, \bar \eta) = r (t) + \eta \; \bar F_1 (t)
+ \bar \eta \; F_1 (t) + i \;  \eta \;  \bar\eta \; q_1 ( t), \nonumber\\
&& \tilde p_{\theta}  (t, \eta, \bar \eta) = p_\theta ( t) + \eta \; \bar F_2 ( t)
+ \bar \eta \; F_2 ( t) + i \; \eta \; \bar\eta \; q_2 ( t), \nonumber\\
&& \tilde \theta (t, \eta, \bar \eta) = \theta (t) + \eta \; \bar F_3 ( t)
+ \bar \eta \; F_3 (t) + i \;  \eta  \; \bar\eta \;  q_3 ( t), \label{sup1}
\end{eqnarray}
and their subsequent substitution  in ({\ref{sup2}}) implies that all the secondary 
variables of the above expansion are zero, namely;
\begin{equation}
F_1 = \bar F_1 = F_2 = \bar F_2 = F_3 = \bar F_3 = q_1 = q_2 = q_3 = 0.
\end{equation}
The above relationships, ultimately, imply the following uniform expansions
\begin{eqnarray} 
&& \tilde r ( t, \eta, \bar \eta) = r (t) + \eta \; (s_{ab} \; r)
+ \bar \eta \; (s_b \; r)  +  \eta \; \bar\eta \; (s_b \; s_{ab} \; r), \nonumber\\
&& \tilde p_{\theta}  (t, \eta, \bar \eta) = p_\theta ( t) + \eta \; (s_{ab} \; p_\theta)
+ \bar \eta \; (s_b \; p_\theta)  +  \eta \; \bar\eta \; (s_b \; s_{ab} \; p_\theta), \nonumber\\
&& \tilde \theta (t, \eta, \bar \eta) = \theta ( t) + \eta \; (s_{ab} \; \theta)
+ \bar \eta \; (s_b \; \theta) +   \eta \; \bar \eta \; (s_b \; s_{ab} \; \theta ).
\end{eqnarray}
The above expansions, finally, lead to $s_{(a)b} \; r = s_{(a)b} \; \theta = 
s_{(a)b} \; p_\theta = 0$. Thus, the trivial (anti-) BRST symmetry transformations
can also be captured within the framework of augmented superfield formalism [11-14].
In our Appendix, we provide the geometrical interpretations of the (anti-) BRST
symmetries (and their corresponding generators) in the language of the translational
generators along the Grassmannian directions of the (1, 2)-dimensional supermanifold. 
Furthermore, we also discuss about the nilpotency and anticommutativity properties
within the framework of superfield formalism. \\

\noindent
{\bf \large 3. (Anti-) co-BRST symmetries: Lagrangian formalism}\\

\noindent
The gauge-fixing term $(\dot \lambda - p_r)$
has its origin in the co-exterior derivative $\delta = \pm * d *$ where
$*$ is the Hodge duality operation on a given manifold. This can be partially 
understood by taking into account $\delta =  * \;d \;*, \; d = dt \; \partial_t$ and
1-form $A^{(1)} = dt \;(\lambda (t))$ on a one-dimensional manifold parametrized
by a single parameter $t$. For instance, it can be checked that
\begin{equation}
\delta A^{(1)} = *\; d \;* \bigl (dt \lambda (t) \bigr ) = \dot \lambda (t), \;\qquad *\;(dt) = 1.
\end{equation}
Thus, the term $\dot \lambda$, in the total gauge-fixing term $(\dot \lambda - p_r)$,
has its origin in the exterior derivative $\delta$. It should be noted that we
have not discussed anything about $p_r$ because it has completely different behavior
under the gauge (or (anti-) BRST) symmetry transformations (cf. (2),(5),(6)).

The following nilpotent ($s_{(a)d}^2=0$) and absolutely 
anticommuting ($s_d s_{ad} + s_{ad} s_d = 0$) 
local infinitesimal transformations $s_{(a)d}$
\begin{eqnarray}
&& s_{ad}\; \lambda = C, \quad s_{ad} \;C = 0, \quad s_{ad}\; p_r = \dot C, 
\quad s_{ad} \;\bar C = - i\; (r - a), \nonumber\\
&& s_{ad}\; b = 0, \qquad s_{ad} \;r = s_{ad} \;\theta = s_{ad} \;p_\theta = 0, 
\qquad s_{ad}\; (\dot \lambda - p_r) = 0, \nonumber\\
&& s_{d} \;\lambda = \bar C, \qquad s_{d}\; \bar C = 0, \qquad s_{d} \;p_r = \dot {\bar C}, 
\qquad s_{d}\; C = i \;(r - a), \nonumber\\
&& s_{d} \;b = 0, \qquad s_{d} \;r = s_{d} \;\theta = s_{d} \;p_\theta = 0, 
\qquad s_{d} \;(\dot \lambda - p_r) = 0, 
\end{eqnarray}
leave the gauge-fixing term invariant\footnote{We came to know about
these transformations from an oral presentation by S. K. Rai much before
his article appeared on the internet (S. K. Rai, B. P. Mandal,
{\it arXiv:1001.5388 [hep-th]}). We differ drastically, however, from
their interpretation of the dual-BRST symmetry transformations as well
as their corroborative logic behind its existence.}. 
As a consequence, we christen them as the
(anti-) dual BRST symmetry transformations. It is straightforward to check
that the FOL (4) remains absolutely invariant under (28) because
\begin{eqnarray}
s_{(a)d} \;L_b = 0.
\end{eqnarray}
Thus, the transformations, listed in (28), are the perfect
symmetry transformations for the first-order BRST invariant Lagrangian (4).

It can be readily checked that the following conserved charges
(that are derived by exploiting the Noether's theorem), namely;
\begin{eqnarray}
Q_d =   b \;\bar C - \dot b \;\dot{\bar C}, \qquad
Q_{ad} =   b \; C - \dot b \;\dot{C},
\end{eqnarray}
are the generators of the transformations (28) because
\begin{eqnarray}
s_r \Psi = - i \;[\Psi, \;Q_r]_{(\pm)}, \qquad r = d, ad, 
\end{eqnarray}
where the subscripts ($\pm$) on the square bracket correspond to the
(anti-) commutator for $\Psi$ being (fermionic) bosonic in nature. It 
is straightforward to check that the following relationships are true, namely;
\begin{eqnarray}
&& s_d Q_d = - i \{ Q_d, Q_d \} = 0, \qquad s_{ad} Q_{ad} = - i \{ Q_{ad}, Q_{ad} \} = 0,
\nonumber\\
&& s_d Q_{ad} = - i \{ Q_{ad}, Q_d \} = 0, \qquad s_{ad} Q_d = - i \{ Q_d, Q_{ad} \} = 0,
\end{eqnarray}
which are basically the reflection of the nilpotency and anticommutativity property of
$s_{(a)d}$ (i.e. $ s_{(a)d}^2 = 0$ and $s_d s_{ad} + s_{ad} s_d = 0 $). The absolute
anticommutativity of $s_d$ and $s_{ad}$, however, imply that only one of them could be
identified with the co-exterior derivative of the differential geometry.

We wrap up this section with the comment that the physicality condition 
$Q_{(a)d} |phys> = 0$ leads to the requirement that $ b |phys> = 0$
and $\dot b |phys> = 0$. These conditions, due to equations (8) and (9),
imply that the operator form of the first-class constraints $\Pi_\lambda 
\approx 0$ and $(r - a) \approx 0$ annihilate the physical states of the 
theory. Thus, the physicality criteria $Q_{(a)d} |phys> = 0$ is consistent 
with the Dirac's method of quantization (see, e.g. [3] for details). It is 
very interesting to point out that, for the present toy model, the (anti-) 
BRST and (anti-) co-BRST charges lead to the {\it same 
conditions} on the physical states due to the requirement of the physicality
criteria ($Q_{(a)b} \;|phys> =0, Q_{(a)d}\; |phys> = 0$)  in the total quantum 
Hilbert space of states (that are consistent with the Dirac's method of quantization).\\

\noindent
{\bf\large 4. Bosonic symmetry: Lagrangian approach}\\

\noindent
It is elementary to check that the following four anticommutators  
\begin{eqnarray}
\{ s_d, s_{ad} \} = 0, \quad \{ s_b, s_{ab} \} = 0,
\quad \{ s_b, s_{ad} \} = 0, \quad \{ s_d, s_{ab} \} = 0,
\end{eqnarray}
are absolutely zero because when they act on any arbitrary dynamical
(and/or auxiliary) variable of the theory,
they produce zero result. The other two anticommutators, constructed from the
four nilpotent ($s_{(a)b}^2 = 0, s_{(a)d}^2 = 0$) operators 
$s_{(a)b}$ and $s_{(a)d}$, are found to be non-zero. These are as follows 
\begin{eqnarray}
\{ s_b, s_d \} = s_{\omega}, \qquad \{ s_{ab}, s_{ad} \} = s_{\bar \omega}.
\end{eqnarray}
The above anticommutators lead to the definition of a unique bosonic symmetry
in the theory. To elucidate this point, let us first express the transformations,
generated by $s_\omega$, as
\begin{eqnarray} 
s_{\omega} \lambda = i \; (b + \dot r), \qquad s_{\omega} p_r 
= i \; (\dot b + r - a),\nonumber\\ 
\qquad s_{\omega} C = s_{\omega} \bar C = s_{\omega} r = s_{\omega} \theta 
= s_{\omega} p_\theta = s_{\omega} b = 0.
\end{eqnarray}
These are symmetry transformations because the FOL remains quasi-invariant as
can be seen from the following explicit expression:
\begin{eqnarray}
s_{\omega} L_b \; = \; \frac{d}{dt} \; \Bigl [i \; (b \dot r - 2 r a + r^2) \Bigr ].
\end{eqnarray}
The action $ S = \int dt \; L_b $, as a consequence, remains invariant if all
the variables of the theory are assumed to fall off rapidly at infinity.

The other anticommutator $\{s_{ad}, s_{ab}\}$ does not lead to an independent 
bosonic symmetry transformation as is evident from the following expression
for the transformations $s_{\bar\omega}$, namely;
\begin{eqnarray}
s_{\bar\omega} \; \lambda \; = \;  - i \; (b + \dot r), \qquad s_{\bar \omega} \; p_r 
\; = \; - i \; (\dot b + r - a),\nonumber\\ 
\qquad s_{\bar \omega} \; C \; = \; s_{\bar \omega} \; \bar C \; = \; s_{\bar \omega} \; r 
\; = \;  s_{\bar \omega} \; \theta \; = \; s_{\bar \omega} \; p_\theta \; = \; 
s_{\bar \omega} \; b \; = \; 0,
\end{eqnarray}
which imply that $ s_{\omega} + s_{\bar \omega} = 0$. As a result we have
the following algebra
\begin{eqnarray}
\{s_b, s_d\} \; = \; s_{\omega} \; = \; - \; \{ s_{ad}, s_{ab} \},
\end{eqnarray}
which demonstrates that $s_\omega$ is the analogue of the Laplacian operator
of differential geometry.  This bosonic symmetry leads  to the following expression
for the conserved charge due to the Noether's theorem:
\begin{eqnarray}
Q_{\omega} \; = \;  i  \; [ b^2 + 2 r a - r^2].
\end{eqnarray}
Using the equations of motion (8), it can be readily checked that
\begin{eqnarray}
\frac{d Q_{\omega}}{dt} \; = i \; \bigl [ 2 b \dot b + 2 \dot r a - 2 \dot r r \;\bigr ] = \; 0. 
\end{eqnarray}
The above conserved charge is the generator of the transformation $s_{\omega}$.
One of the decisive features of the bosonic symmetry is that the (anti-) 
ghost variables of the theory remain unchanged under this transformation. \\

\noindent
{\bf {\large 5. Ghost symmetry: Lagrangian formulation}}\\

\noindent
The ghost number of the (anti-) ghost fields $(\bar C) C$ are $(-1) 1$ and 
the rest of the variables of the theory have ghost number equal to zero.
Thus, we have the following changes of the variables under the ghost-scale transformation:
\begin{eqnarray}
&& r \rightarrow r, \qquad \theta \rightarrow \theta, \qquad p_r \rightarrow p_r,
\qquad p_\theta \rightarrow p_\theta, \nonumber\\
&& \lambda \to \lambda, \quad b \rightarrow b, \quad C \rightarrow e^{\Lambda} C, 
\quad \bar C \rightarrow e^{- \Lambda} \bar C,
\end{eqnarray}
where $\Lambda$ is a global scale parameter. The infinitesimal version of the
above transformations (i.e. $ s_g C = C, \; s_g \bar C = - \bar C, \; s_g \Phi = 0, \; 
\Phi = r, \; \theta, \; p_r, \; p_\theta, \; b,\; \lambda$, etc.) lead to the definition of 
the following conserved charge
\begin{eqnarray}
Q_g = - i\;(\dot {\bar C}\; C + \dot C\; \bar C).
\end{eqnarray}
Using equations of motion, it is pretty easy to prove that $\dot Q_g = 0$. 
This conserved ghost charge, in other words, is the generator of the infinitesimal version of the 
continuous symmetry transformations (41).

In addition to the above continuous symmetry transformation, the ghost sector
respects the following discrete symmetry transformations
\begin{eqnarray}
C \rightarrow \pm \;i \;\bar C, \qquad \bar C \rightarrow \pm\; i\; C.
\end{eqnarray}
The above discrete symmetry transformation is useful in enabling us to
obtain the anti-BRST symmetry transformations from the BRST and {\it vice-versa}.
Furthermore, the above transformations lead to similar kind of relationships
between the co-BRST and the anti-co-BRST symmetry transformations.

We sum up this section with the comment that the (anti-) ghost variables are
decoupled from the rest of the variables of the theory. These variables are unphysical.
As a consequence, the conserved ghost charge does {\it not} put any restriction
on the physical state of the theory. However, the ghost charge does define 
the ghost number of a given state and it plays a major role in establishing
connection between the BRST cohomology and cohomology of the differential forms.
This aspect, we discuss in our next section.\\

\noindent
{\bf \large 6. Cohomological aspects: algebraic structures}\\

\noindent
In this section, we shall establish connection between the conserved charges
(and the continuous symmetry they generate) and the de Rham cohomological
operators. In particular, we shall lay emphasis on the algebraic similarities
between the conserved charges and cohomological operators.\\

\noindent
{\bf\large 6.1 Differential operators and charges}\\

\noindent
It is clear from (9) that all the dynamical variables of the theory have their
corresponding momenta. Thus, the canonical brackets are ($\hbar = c = 1$)
\begin{eqnarray}
&& [r, \;p_r] = i, \qquad [\theta, \;p_\theta ] = i, \qquad [\lambda, \;b] = i, \nonumber\\
&& \{C, \;\dot{\bar C} \} = 1, \qquad \{\bar C, \;\dot C\} = -1,
\end{eqnarray}
and all the rest of brackets are zero. Using these brackets, it can be checked that 
the following algebra is satisfied amongst the conserved charges:
\begin{eqnarray}
&& Q_{(a)b}^2 = 0, \qquad Q_{(a)d}^2 = 0, \qquad \{Q_b, Q_{ab}\} = 0,
\qquad \{Q_d, Q_{ad}\} = 0, \nonumber\\
&& \{Q_b, Q_{ad}\} = 0, \qquad \;\{Q_d, Q_{ab}\} = 0, \;\qquad i\; [Q_g, Q_b] = Q_b, \nonumber\\
&& i\; [Q_g, Q_{ad}] = Q_{ad}, \quad
 i\; [ Q_g, Q_{ab}] = - Q_{ab}, \quad i \;[Q_g, Q_d] = - Q_d, \nonumber\\
&& [Q_{\omega}, Q_r] 
= 0, \qquad \;\;\;\;r = b, \;ab, \;d, \;ad, \;g. 
\end{eqnarray}
This algebra is reminiscent of the algebra satisfied by the de Rham cohomological 
operators of differential geometry. These operators, to be precise, are nothing 
but the exterior derivative $ d $ (with $ d^2 = 0$),
the co-exterior derivative $ \delta = \pm * d *$ (with $\delta^2 =0$ ) and the
Laplacian operator $ \Delta = (d + \delta)^2 = d \delta + \delta d $. Here $*$
is the Hodge duality operation on a given spacetime manifold, on which, the 
above cohomological operators are defined\footnote{ Using directly 
the symmetry transformations (5), (6), (28), (35) and (41) it can be checked that,
in the operator form, they obey the algebra: $ s_{(a)b}^2 = 0, \; s_{(a)d}^2 = 0, 
\; \{s_b, s_{ab}\} = 0, \; \{s_d, s_{ad}\} = 0, \; \{s_b, s_{ad}\} = 0, \;\{s_d, s_{ab}\} = 0,
 \;\ i\; [s_g, s_b] = s_b, \; i\; [s_g, s_{ad}] = s_{ad}, \;
 i\; [ s_g, s_{ab}] = - s_{ab}, \; i \;[s_g, s_d] = - s_d, \; [s_{\omega}, s_r] 
= 0, \;\; r = b, \;ab, \;d, \;ad, \;g. $ Thus, the above symmetry operators also obey the 
algebra of (45) and (46). Hence, they also provide a physical realization of the algebra obeyed by the
cohomological operators of differential geometry.}.

In explicit form, the algebra obeyed by the above de Rham cohomological operators 
of differential geometry are as follows
\begin{eqnarray}
& d^2 = \delta^2 = 0, \quad \Delta = (d + \delta )^2 = d \delta + \delta d 
\equiv \{ d, \delta \}, & \nonumber\\ 
& [ \Delta, \delta ] = 0, \qquad [ \Delta, d ] = 0.&
\end{eqnarray}
Comparing (45) and (46), we obtain following mappings
\begin{eqnarray}
(Q_b, Q_{ad}) \rightarrow d, \quad (Q_d, Q_{ab}) \rightarrow \delta, \quad
Q_{\omega} = \{Q_d, Q_b\} = - \{Q_{ad}, Q_{ab} \} = \Delta,
\end{eqnarray}
which shows that there is two-to-one mapping between the conserved charges 
on one hand and the cohomological operators on the other. Thus, our present
model of rigid rotor provides a toy model for the Hodge theory.

It is well-known that $ d $ raises the degree of a $n$-form $ f_n $ by one
when it acts on it (i.e. $ d f_n \sim f_{n +1}$). On the other hand, the
operator $\delta$ lowers the degree of a $n$-form by one 
when it operates on it (i.e. $ \delta f_n
\sim f_{n - 1} $). The action of the Laplacian 
operators, however, on a $n$-form $ f_n $ keeps the degree intact (i.e. $\Delta
f_n \sim f_n$). These properties are traded with the ghost number of a 
state when we consider the full BRST-cohomology with all the conserved charges.
We discuss below this analogy in a concise manner.

Let $n$ be the ghost number 
of a state $ |\;\psi>_n$ (defined in terms of the ghost charge $Q_g$) 
in the total Hilbert space of states, i.e.,
\begin{eqnarray}
i\; Q_g \; |\; \psi>_n \; = \; n \; |\;\psi>_n,
\end{eqnarray}
then, the following relationships emerge due to the algebra (45), namely;
\begin{eqnarray}
&& i \; Q_g \; Q_b \; |\;\psi>_n \; = \; (n + 1) \; Q_b \; |\; \psi>_n, \nonumber\\
&& i \; Q_g \; Q_{ad} \; |\; \psi>_n \; = \; (n + 1) \; Q_{ad} \;|\; \psi>_n, \nonumber\\
&& i \; Q_g \; Q_d \; |\; \psi>_n \; = \; (n - 1) \; Q_d \; |\; \psi>_n, \nonumber\\ 
&& i \; Q_g \; Q_{ab} \; |\; \psi>_n \; = \; (n - 1) \; Q_{ab} \; |\; \psi>_n, \nonumber\\
&& i \; Q_g \; Q_{\omega} \; |\; \psi>_n \; = \; n  \; Q_{\omega}\; |\; \psi>_n.
\end{eqnarray}
The above equations demonstrate that the ghost number of $ Q_b |\; \psi>_n$ is 
$(n + 1)$, that of $ Q_d|\; \psi>_n $ is $(n - 1)$ and the state $ Q_{\omega} |\; \psi>_n$ 
has the ghost number $n$, respectively. Thus, $(Q_b, Q_d, Q_{\omega})$ form one set that is 
an analogue of $(d, \delta, \Delta)$. Moreover, in an exactly similar fashion,
it can be checked that the other set $(Q_{ad}, Q_{ab}, Q_{\omega})$ also 
obeys the same algebra as $(d, \delta, \Delta)$. Hence, this set also 
constitutes an analogue of the above cohomological operators.

We wrap up this subsection with the conclusion that our present model of rigid
rotor provides a toy model for the Hodge theory where the de Rham cohomological
operators are identified with the Noether's conserved charges (and the continuous 
symmetry transformations they generate). As a consequence, this toy model presents
one of the simplest examples in physics that provides a meeting-ground for some
aspects of differential geometry in mathematics and a few key concepts of 
symmetries in theoretical physics.\\

\noindent
{\bf\large 6.2 Hodge decomposition theorem and conserved charges}\\

\noindent
On a compact manifold without a boundary, it is well-known that an arbitrary 
$n$-form $f_n$ can be written as a unique sum of a harmonic form $h_n$ (i.e. 
$ \Delta h_n = 0 \Rightarrow d h_n = 0, \delta h_n = 0$), an exact form ($ d
e_{n - 1}$) and a co-exact form ($\delta c_{n +1}$) due to the celebrated Hodge decomposition
theorem as [15-17]
\begin{eqnarray}
f_n = h_n + d \;e_{n -1} + \delta \;c_{n +1}.
\end{eqnarray}
The above decomposition can be expressed in the quantum Hilbert space of states
in the following fashion (cf. (47))
\begin{eqnarray}
|\; \psi>_n \; = \; |\; \omega>_n \; + \; Q_b \; |\; \chi>_{n - 1} \; + \; Q_d \; 
|\; \phi>_{n + 1},
\end{eqnarray}
in terms of the set $(Q_b, Q_d, Q_{\omega})$ because $Q_{\omega} |\omega>_n = 0 $ 
implies $Q_b |\omega>_n = 0$ and $Q_d |\omega>_n = 0$. The above equation (51) can
also be expressed in terms of the set $ \bigl ( Q_{ad}, Q_{ab}, Q_{\omega} \bigr )$ 
as follows (cf. (47)) 
\begin{eqnarray}
|\; \psi>_n \; = \; |\; \omega>_n \; + \; Q_{ad} \; |\; \chi>_{n - 1} \; 
+ \;  Q_{ab} \; |\; \phi>_{n + 1},
\end{eqnarray}
where the most symmetric state is the harmonic state $|\;\omega>_n$ that satisfies:
\begin{eqnarray}
Q_{\omega}\;|\; \omega>_n \; = \; 0, \quad Q_{(a)b} \; |\; \omega>_n \; = \; 0, 
\quad Q_{(a)d} \; |\; \omega>_n \; = \; 0.
\end{eqnarray}
In other words, a harmonic state is annihilated by (anti-) BRST as well as (anti-)
co-BRST charges {\it together}. Thus, for aesthetic reasons, it can be chosen as the physical state
because it is the most symmetric state.

It is quite interesting to point out that all the fermionic charges $(Q_{(a)b},
Q_{(a)d})$, due to the following physicality criteria on the physical states, namely;
\begin{eqnarray}
Q_{(a)b} \; |\; phys> = 0, \quad Q_{(a)d} \; |\; phys> = 0, \quad
|phys> = |\mbox {harmonic state}>,
\end{eqnarray}
lead to the following conditions (cf. (8),(9))
\begin{eqnarray}
&& b \;|\; phys> \; = \; 0 \Rightarrow \Pi_{\lambda}\; |\; phys> = 0, \nonumber\\
&& \dot b \;|\; phys> \; = \; 0 \Rightarrow (r - a)\; |\; phys> = 0.
\end{eqnarray}
It is a completely new observation that the full set of all the conserved charges 
($Q_{(a)b}, \; Q_{(a)d}, \; Q_\omega$) of the toy model of the rigid rotor lead 
to the same restrictions on the physical state due to the physicality 
criteria (54). This is true only for the present (very special) toy model.
We note that the operator form of the first-class constraint annihilates 
the physical state as a consequence of the above physicality criteria. 
This outcome is consistent with the Dirac's method of 
quantization of a system with first-class constraints. \\

\noindent
{\bf {\large 7. Conclusions}}\\

\noindent
In our present investigation, we have demonstrated that a toy model
of the rigid rotor, endowed with the first-class constraints in the
language of Dirac's prescription for classification scheme [2,3], is
not only a model for the gauge theory but its (anti-) BRST invariant
version respects a set of six continuous symmetry transformations 
which, in turn, render this theory to be a tractable toy model for
the Hodge theory. As a consequence, the above symmetry transformations
provide a physical realization of the de Rham cohomological operators 
of differential geometry as far as the algebra is concerned.

The BRST symmetry transformation turns out to be the analogue of the exterior 
derivative because the kinetic term, owing its origin to the exterior derivative,
remains invariant under it. In a similar fashion, under the dual(co)-BRST
symmetry transformations, it is the gauge-fixing term (owing partially its origin to
the co-exterior derivative) that remains unchanged. Thus, the co-BRST symmetry
transformation is the analogue of the co-exterior derivative.
The anticommutator of the above two nilpotent symmetries results in a 
bosonic symmetry in the theory which  turns out to be the analogue of
the Laplacian operator. The ghost terms of the theory remain invariant under the 
latter (non-nilpotent) bosonic symmetry transformation.

At the algebraic level, we note that there is two-to-one mapping
between the conserved charges of the theory and the de Rham cohomological
operators (cf. (47)). As a consequence, in the total quantum Hilbert space of
states, there are two ways to express the Hodge decomposition theorem.
The most symmetric state (i.e. the harmonic state) has been chosen,
in our present investigation, as the physical state of the theory.
This state, by its very definition, is annihilated by the four
fermionic (i.e. $Q_{(a)b}^2 = 0, Q_{(a)d}^2 = 0$) charges  and a
bosonic charge ($Q_\omega$) that corresponds to the Laplacian operator.

The physical consequence, that emerges from the restrictions on the
harmonic state with all the conserved charges is, however, one and
the same. In explicit physical terms, this implies that the physical state
(identified with the harmonic state) is annihilated by the operator
form of the first-class constraints of the theory (cf. (55)). This
observation is, however, true only for the present toy model of the rigid
rotor. In the case of the 2D (non-)Abelian 1-form and 4D Abelian 2-form
theories, the restrictions that emerge from the BRST and dual-BRST charges
are different. They are, however, connected by a duality transformation
that is present in these theories (see, e.g. [11-14]). This is not
the case in our present toy model as there is no duality in the (0 + 1)-dimensional
toy model of the rigid rotor.

In our present model, the analogues of the components of the
``gauge'' potential are $\lambda$ and $p_r$. They transform, however,
in a drastically different manner under the gauge and (anti-) BRST
symmetry transformations (cf. (2),(5),(6)). The superfield approach
to BRST formalism (see, e.g. [4,5,11-14]) provides the usual (anti-) BRST
symmetry transformations connected with the usual gauge fields. Thus, it was
a challenge for us to derive the nilpotent (anti-) BRST symmetry transformations
for $\lambda$ and $p_r$ within the framework of geometrical superfield formalism.
It is gratifying to note that we have accomplished this goal in
our subsection 2.2 by exploiting the strength of the modified
version of the HC where we have made some daring (but judicious) choices for the 
(super) fields (see, e.g. (19)). This result might turn
out to be quite useful, later on, in more complex physical situations.

It is interesting to point out that, for the explicit application of the 
idea of HC, we have treated the parameters $r$ and $t$ as independent variables
because, in our view, that is the only judicious option left for us to include both
the components (i.e. $\lambda, p_r$) of the gauge potential {\it together}. Since
the transformation properties of the components of the gauge potential are
radically different (cf.(2),(5),(6)), the latter component (i.e. $p_r$)
has been incorporated (cf. (11),(19)) in a hidden fashion (i.e. $B(r, t) = \partial_r p_r$)
within the framework of superfield formalism and applicability of HC (cf. (15)). Later 
on, we set the limit $r \to 0$. To the best of our
knowledge, our application of HC, in the context of 
the superfield approach to our toy model of rigid rotor, is a novel feature because its 
methodology is quite different from the usual [4,5,11-14].

Our present toy model is one of the simplest models in theoretical
physics that represents a tractable model for the Hodge theory. It
would be a nice future endeavor to look for other toy models (see, e.g. [18,19]) that
could provide examples of the Hodge theory. In fact, such kind
of studies have enabled us to prove that (i) the 2D free
(non-)Abelian gauge theories present a new type of topological
field theories (see, e.g. [6]), and (ii) the 4D free Abelian 2-form
gauge theory is a model for the quasi-topological field theory
(see, e.g. [20]). It would be challenging endeavor to look for
the higher-form and higher-dimensional field theories to provide a set of
tractable field theoretical models for the Hodge theory. To achieve the
above goals, it might be useful to exploit the mathematical
tools (related to the BRST formalism) that are developed in [21,22]. These
issues are under investigation and our results would be
reported in our future publications [23].\\

\noindent
{\bf \large Acknowledgments}\\

\noindent
Financial support from DST, Government of India, under the
SERC project sanction grant No: SR/S2/HEP-23/2006 is gratefully acknowledged.\\

\begin{center}
 {\bf Appendix}
\end{center}

\noindent\\
We discuss here some of the mathematical properties (e.g. nilpotency and 
anticommutativity) associated with the (anti-) BRST symmetry transformations
(and their corresponding generators)
in the language of the geometrical superfield formalism [4,5]. 
From the expansions (22) and (26), it is evident that we have the following 
relationships
\begin{eqnarray}
&& \lim_{\eta \to 0} \; \frac {\partial}{\partial \bar \eta} \; \tilde \Omega^{(h)} \;
(t, \eta, \bar \eta) = s_b \; \Omega (t),  \quad \lim_{ \bar \eta \to 0} \; 
\frac {\partial}{\partial \eta} \; \tilde \Omega^{(h)} \;
(t, \eta, \bar \eta) = s_{ab} \; \Omega (t), \nonumber\\
&& \frac {\partial}{\partial \bar \eta} \; \frac {\partial}{\partial \eta} \; 
\tilde \Omega^{(h)} \; (t, \eta, \bar \eta) = s_b \; s_{ab} \; \Omega (t),
\end{eqnarray} 
where $ \tilde \Omega^{(h)} (t, \eta, \bar \eta)$ is the superfield corresponding 
to the generic dynamical variable $ \Omega (t) \equiv r(t), \; \theta (t), \; p_r (t),\; 
p_\theta (t), \; \lambda (t), \; C (t), \; \bar C (t)$. The former superfields
are obtained after the application of HC and additional
restrictions (cf. (23)). Taking the definition of the generator of a symmetry transformation, it is evident 
that $s_r \Omega (t) = - i \; [ \Omega (t), Q_r ]_{\pm}$ ($r = b, ab$) and the $(+) - $
signs on the square bracket denote (anti-)commutator corresponding to the (fermionic)
bosonic variable $\Omega (t)$. Thus, we have the following mathematical mappings
amongst the various useful (but related) quantities, namely;
\begin{eqnarray}
&& s_b \;  \Longleftrightarrow \; \lim_{\eta \to 0} \;  \frac{\partial}{\partial \bar\eta} 
\; \Longleftrightarrow \; Q_b, \quad  s_{ab} \;  \Longleftrightarrow \; 
\lim_{\bar \eta \to 0} \;  \frac{\partial}{\partial \eta} \; 
\Longleftrightarrow \; Q_{ab}, \nonumber\\
&& s_b \; s_{ab} \;  \Longleftrightarrow  \;  \frac{\partial}{\partial \bar\eta}\;  
\frac{\partial}{\partial \eta} \; \Longleftrightarrow \; Q_b \; Q_{ab},
\end{eqnarray}
which provide the geometrical origin and interpretation for the 
nilpotent and anticommuting (anti-) BRST transformations 
(and their corresponding generators) as the translational operators 
(i.e. $\partial_\eta, \partial_{\bar\eta}$)
along the Grassmannian directions of the (1, 2)-dimensional supermanifold [4,5].

Geometrically, a BRST symmetry transformation $s_b$ (on a variable $\Omega (t)$)
corresponds to the translation of the corresponding superfield $\Omega^{(h)} (t, \eta, \bar\eta)$
along the Grassmannian direction $\bar\eta$ (when there is no translation along the
$\eta$-direction) of the (1, 2)-dimensional supermanifold (on which the present toy
model is generalized). In a similar fashion, an anti-BRST symmetry transformation $s_{ab}$
(on an ordinary variable) is equivalent to the translation of the corresponding superfield
along $\eta$-direction of the (1, 2)-dimensional supermanifold (without any kind
of translation along $\bar\eta$-direction). Two successive translations along any
one of the two Grassmannian directions correspond to the nilpotency property 
present in the BRST formalism. Similarly, the anticommutativity property can be encapsulated
in the statement that the {\it sum} of a translation of the superfield
along $\eta$ (followed by a translations
along $\bar\eta$-direction) {\it and} a translation along $\bar\eta$ (followed by a translation
along $\eta$-direction) results in {\it no} translation at all (i.e. $\partial_\eta \partial_{\bar\eta}
+ \partial_{\bar\eta} \partial_\eta = 0$).

It is quite interesting now to note that the (anti-) BRST charges can be written,
in terms of the superfields, as
\begin{eqnarray}
Q_{ab} & = & i \; \frac {\partial} {\partial \bar \eta} \; \frac {\partial} {\partial \eta}\; 
\Bigl [ \bar F^{(h)} \; \tilde \lambda^{(h)} \Bigr ] \; \equiv \; i \int d \bar \eta 
\; \int d \eta \; \Big ( \bar F^{(h)} \; \tilde \lambda^{(h)} \Bigr ), \nonumber\\
Q_{b} & = & i \; \frac {\partial} {\partial \bar \eta} \; \frac {\partial} {\partial \eta}\; 
\Bigl [ F^{(h)} \; \tilde \lambda^{(h)} \Bigr ] \; \equiv \; i \int d \bar \eta 
\; \int d \eta \; \Big ( F^{(h)} \; \tilde \lambda^{(h)} \Bigr ),
\end{eqnarray}
where superscript $(h)$ denotes the superfields obtained after the application 
of HC (cf. (21), (22)). Written in terms of the nilpotent (anti-) BRST symmetry transformations
(cf. (5), (6)), the above equations read as
\begin{eqnarray}
Q_{ab} = i \; s_b \; s_{ab} \; (\bar C \; \lambda) \; \equiv \; - i \; s_{ab} 
\; s_b \; ( \bar C \; \lambda), \nonumber\\
Q_{b} = i \; s_b \; s_{ab} \; (C \; \lambda) \; \equiv \; - i \; s_{ab} 
\; s_b \; (C \; \lambda).
\end{eqnarray}
This establishes the nilpotency and anticommutativity of the (anti-) BRST symmetries 
(and their corresponding generators) as is evident from the following useful relationships:
\begin{eqnarray}
s_b \; Q_{ab} &=& - i \; \{ Q_{ab},\;  Q_b\} = 0, \quad  
s_b \; Q_b = - i \; \{ Q_b,\;  Q_b\} = 0, \nonumber\\
s_{ab} \; Q_{ab} &=& - i \; \{Q_{ab},\;  Q_{ab}\} = 0, \quad
s_{ab} \; Q_b = - i \; \{Q_b,\;  Q_{ab}\} = 0.
\end{eqnarray}
In the above, we have exploited the definition of the generator 
(cf., e.g. (31)) of a continuous symmetry transformation.

The above BRST charge $Q_b$  can be expressed in two more different ways. 
These, expressed in terms of the (anti-) ghost and gauge superfields, are 
\begin{eqnarray}
 Q_b  &=&   - i  \lim_{\eta \to 0} \frac{\partial}{\partial \bar \eta} \Bigl [ F^{(h)}  \dot {\bar F}^{(h)}
+ i  b \tilde \lambda^{(h)} \Bigr ] \;
 \equiv  - i  \lim_{\eta \to 0} \int d \bar \eta  
\Bigl [ F^{(h)} \dot {\bar F}^{(h)} + i b \tilde \lambda^{(h)} \Bigr ],\nonumber\\
 Q_b \; &=&  \; i \; \frac{\partial}{\partial \eta}\; \bigl(F^{(h)} \dot F^{(h)} \bigr)
\;  \equiv \; i \int d \eta \; \bigl( F^{(h)} \dot F^{(h)} \bigr), 
\end{eqnarray}
which can be re-expressed in terms of the nilpotent (anti-) BRST symmetry transformations 
(cf. (5), (6)) as
\begin{eqnarray}
Q_b = i \; s_{ab} \; ( C \; \dot C), \qquad  Q_b = - i \; s_b \; 
[ C \; \dot {\bar C} + i \; b \; \lambda ].
\end{eqnarray}
Once again, the nilpotency and anticommutativity properties become quite 
transparent from (61) and (62). Furthermore, it is clear that, for this simple
toy model, the (anti-) BRST charges are BRST as well as anti-BRST invariant.
This can be proven by exploiting the nilpotency $s_{(a)b}^2 = 0$ property.

In exactly similar fashion, the anti-BRST charge ($Q_{ab}$) can be expressed,
in two different and distinct ways, as listed below:
\begin{eqnarray}
&& Q_{ab} = - i \; \frac {\partial}{\partial \bar \eta} \; \Bigl [ \bar F^{(h)} 
\; \dot {\bar F}^{(h)} \Bigr ] \; \equiv \; - i \; s_b \;  
\bigl ( \bar C \; \dot {\bar C} \bigr ),\nonumber\\
&& Q_{ab} = - i \; \lim_{\bar \eta \to 0}\; \frac {\partial}{\partial \eta} \; \Bigl [ \dot F^{(h)} \bar F^{(h)} 
+ i \; b\; \tilde \lambda^{(h)} \Bigr ] \equiv  - i  s_{ab} \;  \bigl ( \dot C  \bar C +
i b \lambda \bigr ).
\end{eqnarray}
The above expressions can also be written in terms of the integrals over the Grassmannian 
variables as follows
\begin{eqnarray}
Q_{ab} = - i \; \int d \bar \eta \; \Bigl [ \bar F^{(h)} \; \dot {\bar F}^{(h)} \Bigr ]
\; \equiv \; - i \; \int d \eta \; \Bigl [ \dot F^{(h)} \; \bar F^{(h)} + 
i\; b \; \tilde \lambda ^{(h)} \Bigr ].
\end{eqnarray} 
In the above form, too, the nilpotency of the anti-BRST charge ($Q_{ab}$) and its 
anticommutativity property with the BRST charge ($Q_b$) become quite lucid.

We close this Appendix with the remark that there are alternative ways to express the nilpotent
(anti-) BRST charges in the superfield formulation. It can be checked that, in equations
(58), (61), (63) and (64), one can replace $\tilde \lambda^{(h)} \to \dot {\tilde p}_r^{(h)}$ (cf. (21))
without changing the algebraic expressions. Furthermore, the nilpotency and anticommutatvity
properties are encoded in the properties of the translational generators $\partial_\eta$
and $\partial_{\bar\eta}$. For instance, it can be checked that the observations:
$\partial_\eta Q_{(a)b} =0$ and
$\partial_{\bar\eta} Q_{(a)b} = 0$, imply the above two properties due to (56). Thus, the 
superfield formalism does provide the geometrical meaning of the nilpotency and 
anticommutativity properties. \\

\end{document}